\documentclass{PoS}

\usepackage[utf8x]{inputenc}

\usepackage{mciteplus}

\newcommand{\gev}{\ \textrm{GeV}}

\newcommand{\lqcd}{\Lambda_{\mathrm{QCD}}}
\newcommand{\as}{\alpha_{\mathrm{s}}}

\newcommand{\rt}{{\mathbf{r}}}
\newcommand{\xt}{{\mathbf{x}}}
\newcommand{\bt}{{\mathbf{b}}}

\newcommand{\yt}{{\mathbf{y}}}
\newcommand{\zt}{{\mathbf{z}}}

\newcommand{\kt}{{\mathbf{k}}}

\newcommand{\tr}{\, \mathrm{Tr} \, }
\newcommand{\nc}{{N_\mathrm{c}}}

\newcommand{\der}{\mathrm{d}}

\newcommand{\xpom}{{x_\mathbb{P}}}

\newcommand{\A}{{\mathcal{A}}}

\title{Impact parameter dependent JIMWLK evolution meets HERA data}

\ShortTitle{JIMWLK meets HERA}

\author{\speaker{Heikki Mäntysaari}\\
        Department of Physics, University of Jyväskylä, P.O. Box 35, 40014 University of Jyväskylä, Finland\\
        E-mail: \email{heikki.mantysaari@jyu.fi}}

\author{Björn Schenke\\
        Physics Department, Brookhaven National Laboratory, Upton, NY 11973, USA\\
        E-mail: \email{bschenke@bnl.gov}}

\abstract{
We calculate the small-$x$ evolution of protons with finite size by solving numerically the JIMWLK evolution equation. The initial condition is constrained by the HERA measurements of charm reduced cross section and of exclusive vector meson production. We compute the energy dependence of diffractive cross sections in order to access the energy evolution of the event-by-event fluctuating proton density profile. The fundamental problems arising from the regime sensitive to non-perturbatively large dipoles are also discussed.
}

\FullConference{International Conference on Hard and Electromagnetic Probes of High-Energy Nuclear Collisions\\
		30 September - 5 October 2018\\
		Aix-Les-Bains, Savoie, France}

\begin{document}

\section{Introduction}

Thanks to its pointlike structure, a high energy electron is a clean probe of the internal structure of a hadron. To date, the most detailed measurements that probe the partonic substructure of the proton have been performed by the HERA experiments H1 and ZEUS. For example, the accurately measured structure functions~\cite{Abramowicz:2015mha,H1:2018flt} are proportional to the parton densities and make it possible to extract the quark and gluon distributions over a wide range of longitudinal momentum fraction $x$ and virtuality $Q^2$. On top of that, additional information about the transverse geometry can be obtained from studies of exclusive vector meson production also studied extensively at HERA (see e.g.~\cite{Chekanov:2002xi,Aktas:2005xu,Alexa:2013xxa}).

In exclusive (or diffractive) vector meson production one can measure the total momentum transfer, which is Fourier conjugate to the impact parameter. Consequently, the transverse density of the partonic content can be extracted. Additionally, in case of incoherent diffraction where the proton dissociates, it becomes possible to access not only the transverse density profile, but also its event-by-event fluctuations~\cite{Miettinen:1978jb, Mantysaari:2016ykx, Mantysaari:2016jaz}.

In the future, the Electron Ion Collider~\cite{Accardi:2012qut}, complemented by ultraperipheral heavy ion collisions studied at CERN,  will provide precise measurements of the proton (and nuclear) structure over a wide kinematical range, making it possible to study non-linear QCD effects and to constrain the initial state of heavy ion collisions. In this context, we have studied in Ref.~\cite{Mantysaari:2018zdd} the energy evolution of the gluonic structure of the proton at small $x$.

\section{Structure functions and vector meson production}

The two processes considered in this work are inclusive photon-proton scattering (proportional to the reduced cross section) and diffractive vector meson production. At high energies, these processes are conveniently studied in the dipole picture, where the incoming virtual photon splits into a quark-antiquark pair long before the interaction with the proton takes place. The quarks are color rotated in the proton color field picking up Wilson lines, and finally they form the vector meson. In case of inclusive processes, the optical theorem relates the forward elastic scattering amplitude $\gamma^* + p \to \gamma^* + p$ to the total inelastic cross section. A necessary ingredient in both cases is the dipole-proton scattering amplitude $N(\rt,\bt,x)$ for the dipole with transverse size $\rt$ and impact parameter $\bt$. In terms of Wilson lines, which contain all the information about the target, the dipole amplitude is written as $1- N(\xt, \yt, x) = \tr V(\xt)V^\dagger(\yt) / \nc$ with $\rt=\xt-\yt$ and $\bt=(\xt+\yt)/2$. 

In case of diffractive vector meson production, the cross section is a convolution of the photon and vector meson ($V$) wave functions (describing processes $\gamma \to q\bar q$ and $q\bar q\to V$), Fourier transformed from transverse coordinate space to transverse momentum space. The scattering amplitude reads (see e.g.~\cite{Kowalski:2006hc})
\begin{equation}
\label{eq:diff_amp}
 \A^{\gamma^* p \to V p}_{T,L}(\xpom,Q^2, { \Delta})= 2i\int \der^2 \rt \int \der^2 \bt \int \frac{\der z}{4\pi}   (\Psi^*\Psi_V)_{T,L}(Q^2, \rt,z) e^{-i[\bt - (1-z)\rt]\cdot { \Delta}}  N(\rt,\bt,\xpom).
\end{equation}
Here $\Psi^*\Psi_V$ denotes the overlap between the virtual photon and the vector meson wave functions (see ~\cite{Kowalski:2006hc} for details)
 and $|\Delta|=\sqrt{-t}$ is the transverse momentum of the vector meson.  The longitudinal momentum fraction of the photon carried by the quark is denoted by $z$, and $T,L$ refer to the transverse and longitudinal photon polarization, respectively.

 In case of coherent scattering, in which the target remains intact, the cross section is $\der \sigma / \der t = |\langle  \A^{\gamma^* p \to V p}_{T,L} \rangle |^2/(16\pi)$. The average over target configurations is denoted by  $\langle \rangle$. In the incoherent scattering the cross section becomes a variance: $\der \sigma / \der t = \left[ \langle |  \A^{\gamma^* p \to V p}_{T,L} |^2 \rangle -   |\langle  \A^{\gamma^* p \to V p}_{T,L} \rangle |^2 \right]/(16\pi)$. As a variance, it measures the amount of event-by-event fluctuations in the scattering amplitude (and  in the impact parameter profile of $N$), whereas the coherent cross section is sensitive to the average shape of the target. 
The total cross section for the $\gamma^*p$ scattering is obtained by applying the optical theorem:
\begin{equation}
\label{eq:gamma-xs}
	\sigma^{\gamma^* p}_{L,T} = 2 \sum_f \int \der^2 \bt \der^2 \rt \frac{\der z}{4\pi} \left|\Psi^f_{L,T}(r,z,Q^2)\right|^2 \langle N(\rt, \bt, x) \rangle,
\end{equation}
where $\langle N(\rt, \bt, x) \rangle$ is the average dipole-target scattering amplitude. The sum $f$ runs over quark flavors.

The Wilson lines are obtained as follows\footnote{A similar setup is used in Ref.~\cite{Schlichting:2014ipa}}. The initial condition for the JIMWLK evolution is determined by applying the MV model~\cite{McLerran:1993ni} . The color charge density at each point in the transverse plane is a local random gaussian variable with correlator proportional to $g^4\mu^2T_p(\bt)$.  In case we do not include proton substructure, the transverse profile $T_p$ is a simple Gaussian $T_p(\bt)=e^{-b^2/(2B_p)}$, where the proton size at initial $x_0$ is controlled by $B_p$. When we also want to study the event-by-event fluctuations, $T_p$ becomes a sum of $N_q$ Gaussians located around the sampled hot spot positions. The Wilson lines are obtained by solving the classical Yang-Mills equations. Here, we also study an initial condition where we modify the MV model by introducing an additional ultraviolet suppression factor, by replacing in  momentum space $A^+(x^-, \kt) \to A^+(x^-, \kt)e^{-|\kt| v}$.

The energy evolution of the Wilson lines is obtained by solving the JIMLWK renormalization group equation. In the Langevin form, which is suitable for numerical calculations, it can be written as
$
\frac{\der}{\der y} V_\xt = V_\xt (i t^a) \left[
\int \der^2 \zt\,
\varepsilon_{\xt,\zt}^{ab,i} \; \xi_\zt(y)^b_i  + \sigma_\xt^a 
\right].
$
Here $\xi$ is a random Gaussian noise with coefficient $\varepsilon$, and $\sigma$ is a deterministic drift term. For detailed expressions and details about the implementation, the reader is referred to Ref.~\cite{Mantysaari:2018zdd} and references therein.

\section{Results}
The free parameters in our framework are the proton density $g^4\mu^2$, its size $B_p$ (and possible substructure fluctuations) and the value of the strong coupling $\as$ (or coordinate space $\lqcd$ when we use running coupling) in the JIMWLK evolution. Additionally, we have to introduce infrared regulators to suppress long distance Coulomb tails when sampling the initial condition and also in the JIMWLK evolution. The parameters are fixed by fitting the HERA combined measurement of the charm structure function data~\cite{H1:2018flt} as it is not sensitive to large dipole sizes~\cite{Mantysaari:2018nng}.

\begin{figure*}[tb]
\centering
\begin{minipage}{0.48\textwidth}
		\includegraphics[width=\textwidth]{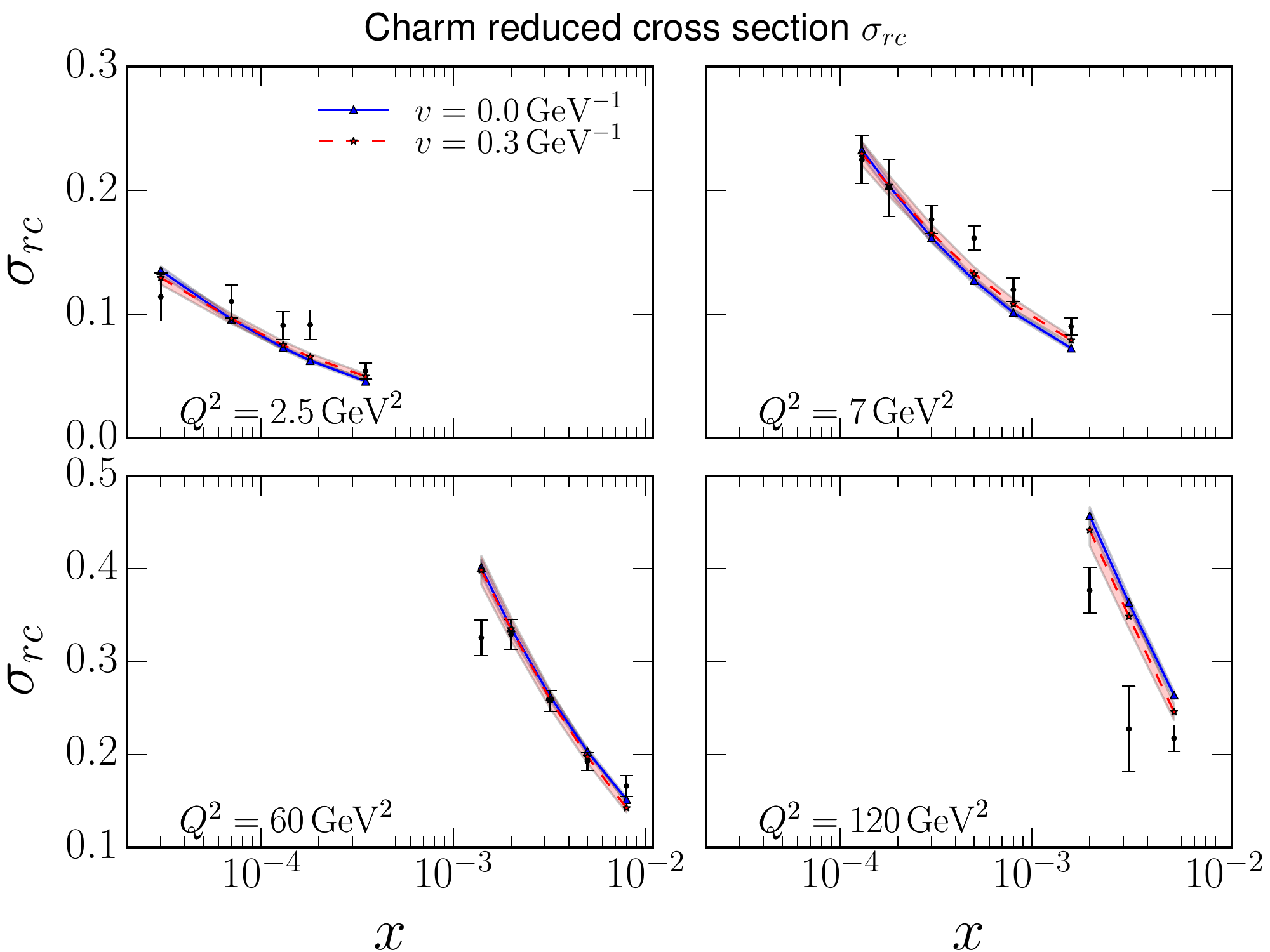} 
				\caption{Description of the HERA charm production data~\cite{H1:2018flt}. Figure from Ref.~\cite{Mantysaari:2018zdd}.		}
		\label{fig:sigmar}
\end{minipage}
\quad
\begin{minipage}{0.48\textwidth}
\centering
		\includegraphics[width=\textwidth]{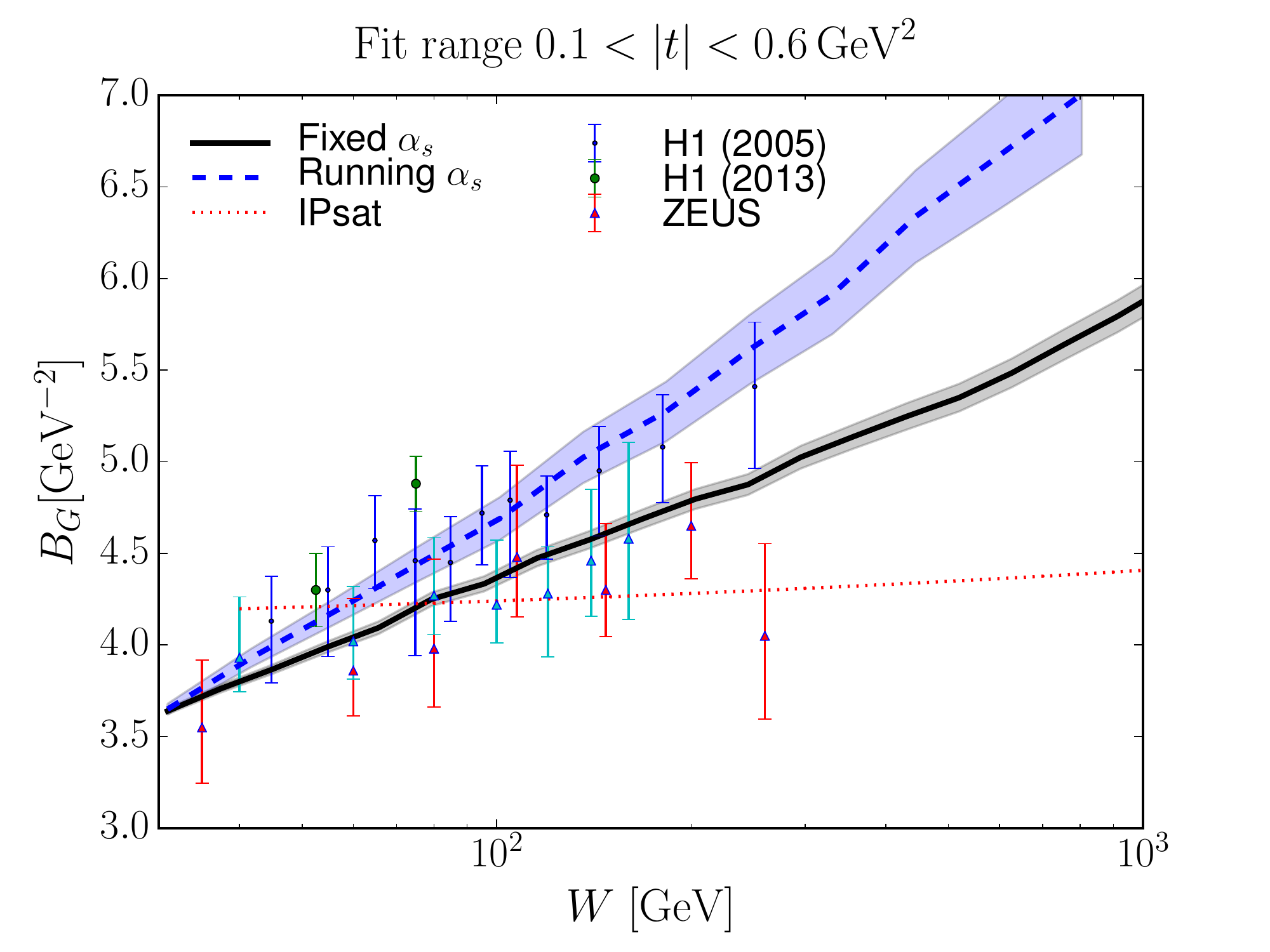} 
				\caption{Slope of the diffractive $J/\Psi$ production compared with the HERA data~\cite{Chekanov:2002xi,Aktas:2005xu,Alexa:2013xxa}. Figure from Ref.~\cite{Mantysaari:2018zdd}.}
				
		\label{fig:slope_mdep_mint_0.1}
\end{minipage}
\end{figure*}

The description of the charm reduced cross section data  using a proton with no hot spot substructure is shown in Fig.~\ref{fig:sigmar}. In the figure the results with fixed coupling JIWMLK evolution are shown, the result at running coupling being identical. With an ultraviolet suppression factor $v>0$, the $Q^2$ evolution is better described and $\chi^2/N$ improves.

The proton size is extracted as the $t$ slope of the coherent exclusive $J/\Psi$ production cross section. The results are shown in  Fig.~\ref{fig:slope_mdep_mint_0.1}. With running coupling the proton grows  faster than at fixed coupling. This is due to the fact that in both cases the average evolution speed is similar (describes $\sigma_r$), but at running coupling the evolution of the short-wavelength modes is suppressed relative to long wavelength ones that dominate the proton size measure at low $|t|$.

\begin{figure*}[tb]
 \begin{minipage}{0.48\textwidth}
\centering
		\includegraphics[width=\textwidth]{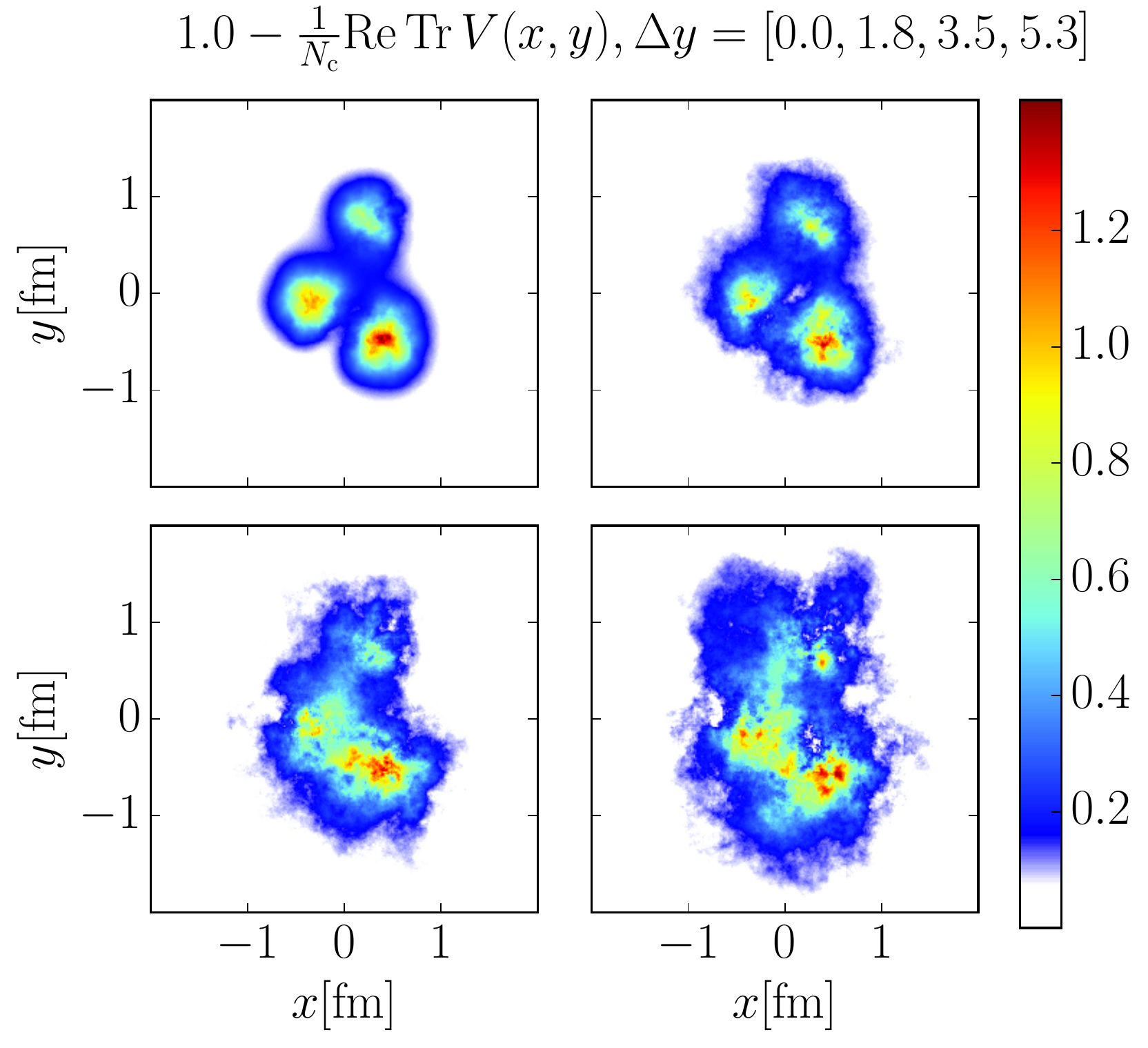} 
				\caption{Illustration for the evolution of the proton density with substructure and $v=0$. Figure from~\cite{Mantysaari:2016ykx}.}
				
		\label{fig:proton_wlines}
\end{minipage}
\quad
\centering
\begin{minipage}{0.48\textwidth}
		\includegraphics[width=\textwidth]{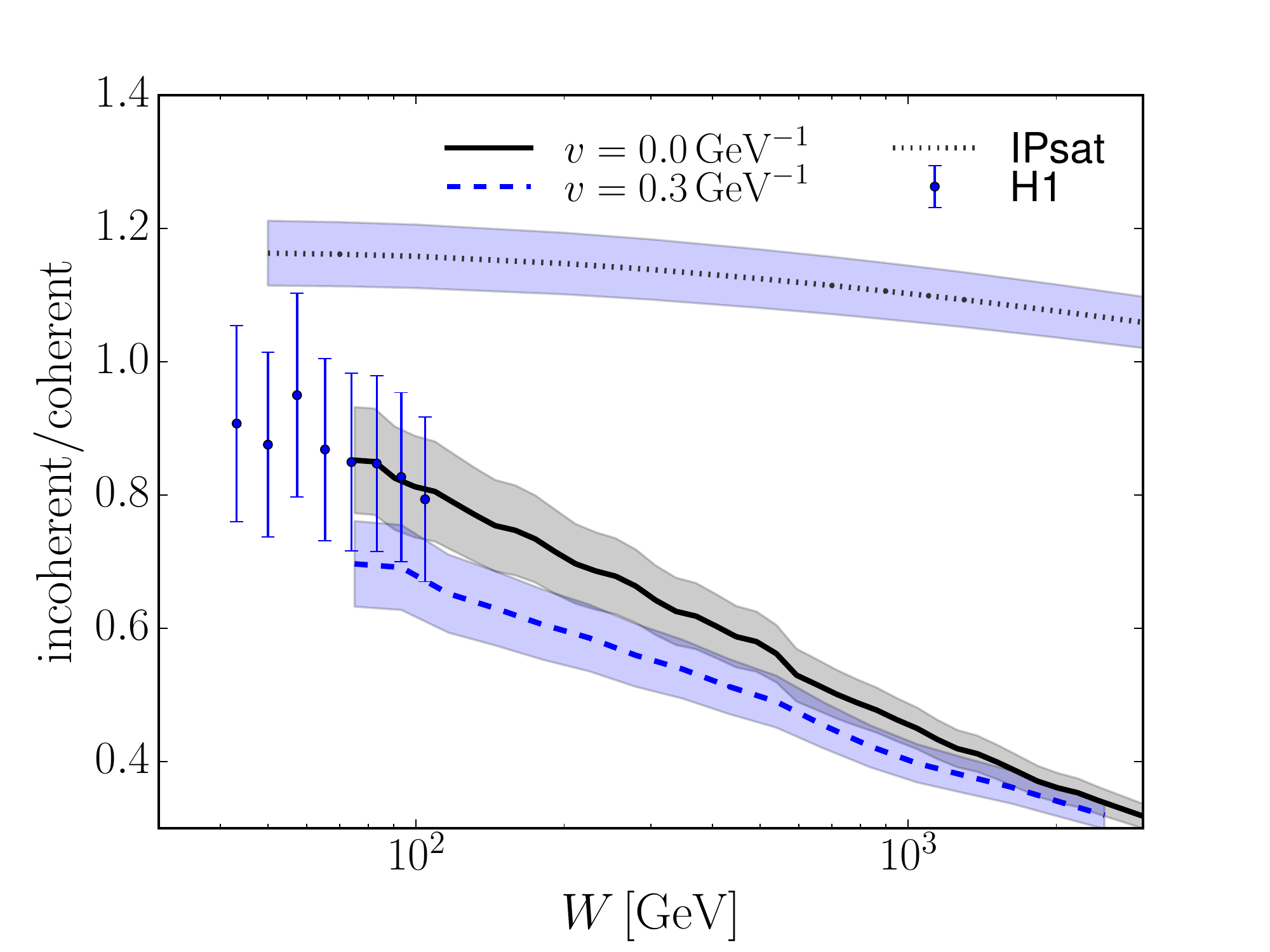} 
				\caption{Ratio of incoherent to coherent cross section. The experimental uncertainties are computed by assuming independent errors for total coherent and incoherent cross sections~\cite{Alexa:2013xxa}. Figure from~\cite{Mantysaari:2016ykx}. 	}
		\label{fig:incohcohratio}
\end{minipage}
\quad
\end{figure*}

To study the evolution of the proton geometry, we use a proton density profile $T_p$ that consists of $N_q=3$ hot spots and fix parametrization such that the HERA coherent and incoherent $J/\Psi$ measurements at  $W=75\gev$ (corresponding to $x\approx 10^{-3}$) are reproduced, similarly as in Ref.~\cite{Mantysaari:2016ykx}. Then, we use the same JIMWLK evolution constrained by the charm $\sigma_r$ data, and compute diffractive cross sections at high energies. The resulting evolution of the proton density is illustrated in Fig.~\ref{fig:proton_wlines} for a particular configuration. In Fig.~\ref{fig:incohcohratio} we show the obtained ratio for the incoherent and coherent cross sections compared with the H1 data. For comparison, in Fig.~\ref{fig:incohcohratio} the result from an IPsat model calculation, where there is no geometry evolution, is shown. In that case the proton does not get less lumpy at small $x$, and the cross section ratio is practically flat.

When applying the framework to describe the total structure function with light quark contribution included, we find that the obtained results significantly underestimate the HERA data. Similarly, the fit that describes charm $\sigma_r$ data underestimates the $J/\Psi$ production cross section. The reason is that in our calculation dipoles where both quarks miss the proton do not contribute. This is probably unrealistic, and one should include a separate description for the confinement scale effects affecting these dipoles. One such a description is studied in Ref.~\cite{Mantysaari:2018zdd}.

\section{Conclusions}

The initial condition to the JIMWLK evolution is fitted to the HERA charm production data. 
 The resulting evolution of the proton structure is compatible with the measurement of the (gluonic) size of the proton measured in exclusive $J/\Psi$ production, but inclusive structure function and normalization of the  $J/\Psi$ production cross section are underestimated. By calculating event-by-event evolution for the fluctuating proton structure, we find that the initial hot spot structure is washed out and the resulting energy dependence of the cross section ratio is compatible with the H1 data.

\section*{Acknowledgments}
H.M. is supported by the Academy of Finland, project 314764, and by the European Research Council, Grant ERC-2015-CoG-681707.  BPS is supported under DOE Contract No. DE-SC0012704 and acknowledges a DOE Office of Science Early Career Award.
 Computing resources of the National Energy Research Scientific Computing Center supported by the Office of Science of the U.S. Department of Energy under Contract No. DE-AC02-05CH11231 and of the CSC -- IT Center for Science in Espoo, Finland were used.
\bibliographystyle{h-physrev4mod2}
\bibliography{../../refs}

\end{document}